\documentclass[%
 aip,
 amsmath,amssymb,
 reprint,%
]{revtex4-1}

\usepackage{graphicx}
\usepackage{dcolumn}
\usepackage{bm}

\usepackage[utf8]{inputenc}
\usepackage[T1]{fontenc}
\usepackage{mathptmx}
\usepackage{etoolbox}
\usepackage{siunitx}

\makeatletter
\def\@email#1#2{%
 \endgroup
 \patchcmd{\titleblock@produce}
  {\frontmatter@RRAPformat}
  {\frontmatter@RRAPformat{\produce@RRAP{*#1\href{mailto:#2}{#2}}}\frontmatter@RRAPformat}
  {}{}
}%
\makeatother
\begin{document}

\preprint{AIP/123-QED}

\title{Low dephasing and robust micromagnet designs for silicon spin qubits}
\author{N. I. Dumoulin Stuyck}
 \affiliation{Department of Materials Engineering (MTM), KU Leuven, B-3001 Leuven, Belgium}
 \affiliation{imec, B-3001 Leuven, Belgium}
\author{F. A. Mohiyaddin}%
  \affiliation{imec, B-3001 Leuven, Belgium}
\author{R. Li}
  \affiliation{imec, B-3001 Leuven, Belgium}
  \email{roy.li@imec.be}
\author{M. Heyns}
 \affiliation{Department of Materials Engineering (MTM), KU Leuven, B-3001 Leuven, Belgium}
  \affiliation{imec, B-3001 Leuven, Belgium}
\author{B. Govoreanu}
  \affiliation{imec, B-3001 Leuven, Belgium}
\author{I. P. Radu}
  \affiliation{imec, B-3001 Leuven, Belgium}
  
\date{\today}

\begin{abstract}
Using micromagnets to enable electron spin manipulation in silicon qubits has emerged as a very popular method, enabling single-qubit gate fidelities larger than $99.9\%$. However, these micromagnets also apply stray magnetic field gradients onto the qubits, making the spin states susceptible to electric field noise and limiting their coherence times. We describe here a magnet design that minimizes qubit dephasing, while allowing for fast qubit control and addressability. Specifically, we design and optimize magnet dimensions and position relative to the quantum dots, minimizing dephasing from magnetic field gradients. The micromagnet-induced dephasing rates with this design are up to 3-orders of magnitude lower than state-of-the-art implementations, allowing for long coherence times. This design is robust against fabrication errors, and can be combined with a wide variety of silicon qubit device geometries, thereby allowing exploration of coherence limiting factors and novel upscaling approaches.
\end{abstract}

\maketitle

Silicon quantum dot-based spin qubit systems have demonstrated attractive properties for quantum computation, including long qubit coherence times and high-fidelity single- and two-qubit gates\cite{Veldhorst2014,Veldhorst2015,Yoneda2018,Yang2019,Huang2019}. 
The introduction of patterned on-chip micromagnets has allowed high-speed and full electrical spin manipulation using Electric Dipole Spin Resonance (EDSR) \cite{NadjPerge2010,Tokura2006,Watson2018,Noiri2016,Kawakami2014,Takakura2010,Struck2020,Takeda2020,Pioro-Ladriere2007,Zajac2018}, and enabled $>99.9\%$ single-qubit gate fidelity so far\cite{Yoneda2018}. Furthermore, micromagnets have sparked novel physical mechanisms for spin control, read-out and manipulation \cite{Mi2017,Samkharadze2018}. However, the micromagnet generated gradients are a double-edged sword, as detrimental magnetic field gradients in combination with electric field noise result in substantial dephasing of the qubit. Indeed, successful demonstrations of EDSR with spin qubits have reported charge noise as the limiting factor for longer qubit coherence times and higher spin control fidelities\cite{Yoneda2018,Struck2020,Takeda2020}. The open challenge is to minimize the qubit dephasing rate while still allowing fast manipulation and qubit addressability in larger-scale arrays. In this Letter, we present optimized micromagnet designs that address this challenge and facilitate both low qubit dephasing and fast control, as well as ensure individual qubit addressability for large quantum dot array implementations. In comparison to previous works, which mainly focused on qubit manipulation speed and addressability\cite{Yoneda2015,osti2015,Zhang2021}, here we use multiphysics based modelling that correlate design parameters to the qubit performance, and co-optimize the dephasing field gradients. The presented design and insight can be readily and widely implemented in popular Si-MOS and Si/SiGe based spin qubit architectures. By resolving the micromagnet induced dephasing, the design will allow further exploration of coherence-limiting factors, crucial for the development of large-scale spin qubit arrays\cite{Li2018,Vandersypen2017}. 
\\ \\
The remainder of the paper is organised as follows. We start by introducing a popular micromagnet topology along with its relevant design parameters, while stressing that the optimization is extendable to other micromagnet device topologies. We show the typical magnetic fields and their gradients that can be expected from the design. Next, we demonstrate an optimal design point with minimal dephasing field gradients that is robust against misplacement errors due to fabrication. We show that, at this optimal point, low dephasing can be achieved while maintaining fast qubit manipulation speeds and qubit addressability. Finally, we estimate expected micromagnet-induced dephasing rates and coherence times in the presence of charge noise, highlighting an improvement by over three orders of magnitude in our optimized design.

\begin{figure}
\includegraphics{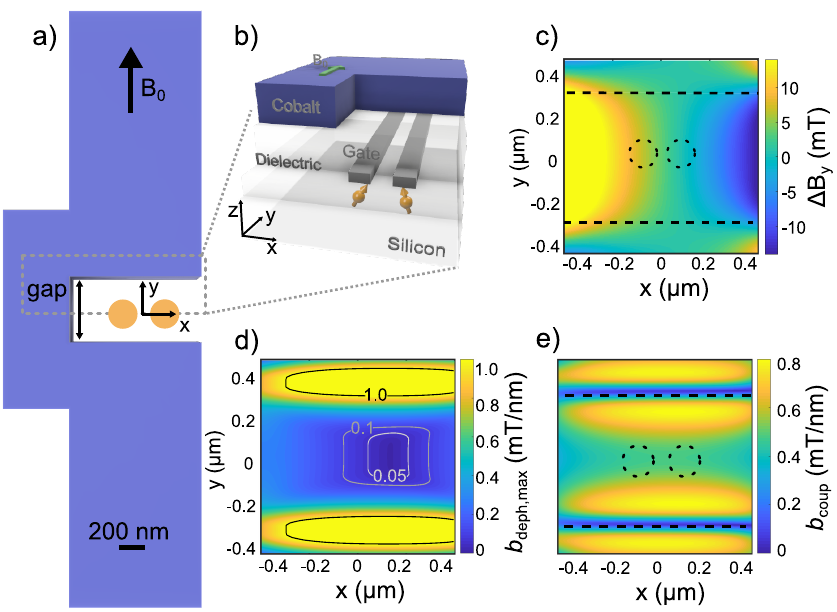}
\caption{\label{fig:1b} a) Design of the micromagnet. The x-y-z coordinate axis denotes the relative position of the micromagnet to the gate defined quantum dots, shown in b). For simplicity, only two quantum dots are shown but this number can be extended to a larger array. A magnetizing field $B_0$ is applied along the y-direction and splits the confined electron’s spin states. c) Magnetic field difference along the quantum dot placements (x-direction) which allows for qubit addressability. d) Maximum of dephasing gradients $b_{\text{deph}}^{\text{dy}}$ and $b_{\text{deph}}^{\text{dx}}$. e) coupling field gradient generated by the magnets. Fields are simulated for a $\SI{600}{\nano \meter}$ gap design and taken at $d_{\text{QDM}} = \SI{120}{\nano \meter}$. Dashed circles denote typical quantum dot locations, horizontal dashed lines show the micromagnet outline.}
\end{figure}

The qubit states are encoded in the spin states of an electron confined in a gated nanostructure \cite{Loss1997}, as shown in Fig.~\ref{fig:1b}(b). Spin manipulation via EDSR is performed by displacing the electron wavefunction through gate generated electric fields in magnetic field gradients, locally generated by micromagnets. The micromagnet design is shown in Fig.~\ref{fig:1b}(a), as well as its relevant dimensions, design parameters and chosen axis origin. For typical EDSR experiments, qubits are placed in the center of the gap along the x-direction to achieve high manipulation speed and addressability \cite{Yoneda2015}. We study the magnetization of the magnet and the resulting 3D magnetic fields by solving the Landau-Lifshitz-Gilbert equation numerically\cite{Donahue1999,Mohiyaddin2019}, and we refer to the supplementary material for all spatial components and gradients of the magnetic field. We chose a numerical approach to be able to accurately determine the magnetic fields for non-regular micromagnets, which might not be fully magnetized at smaller magnetic fields. Here, we chose Cobalt as the micromagnet material because of its strong ferromagnetism and widespread use in spin-based quantum computation\cite{Yoneda2015,Noiri2016,Kawakami2014,Struck2020,Watson2018,Tokura2006,Shin2010,Mi2017,Samkharadze2018}. Furthermore, we focus on a magnet thicknesses of $\SI{200}{\nano \meter}$ as it is a commonly used value, and has a relatively weak influence on the magnetic fields. For the external applied magnetic field, typical values are in the range of a few $\SI{100}{\milli \tesla}$ to $\sim$1 T\cite{Yoneda2015,Noiri2016,Kawakami2014,Struck2020,Watson2018,Tokura2006}. We studied the effect of this external field on the magnetization and found the micromagnet optimization depends only weakly on magnetization fields changing between $\SI{100}{\milli \tesla}$ and $\SI{10}{\tesla}$ (see supplementary material for details). We use a $B_0$ of 0.7 T along the y-direction in the following results and discussions, and refer the reader to the supplementary material for the full system Hamiltonian.
In this architecture, the qubit’s coherent two-level system is formed by the electron spin up and down state, which are split in energy by the external applied magnetic field $B_0$ and micromagnet field $B_y$ along the y-direction. This energy splitting is given by the Zeeman energy $E_Z = \gamma_e B_0$, where $\gamma_e \approx \SI{116}{\micro \electronvolt / \tesla}$ is the electron gyromagnetic ratio. To identify optimal magnet dimensions, we define the following parameters:
\begin{align*}
b_{\text{deph}}^{\text{dx,dy}}(x,y) &= \max_{X,Y \in A(x,y) } \left[ \left|\frac{dB_y}{dx,dy}(X,Y) \right| \right] \\
b_{\text{coup}}(x,y) &=\min_{X,Y \in A(x,y)} \left[\sqrt{\left(\frac{dB_x}{dy}(X,Y) \right)^2 + \left(\frac{dB_z}{dy}(X,Y) \right)^2}\right] \\
\Delta B_y(x,y) &=\left|B_y (x -  \SI{50}{\nano \meter},y) - B_y (x + \SI{50}{\nano \meter},y)\right|, \\
\end{align*}
with $A(x,y)$ defined as the quantum dot footprint of $50 \times \SI{50}{\nm^2}$, centred around $(x,y)$. Taking the maximum of $b_{\text{deph}}$ and minimum of $b_{\text{coup}}$ over $A(x,y)$ ensures the worst-case scenario in electron wavefunction distribution, taking into account the quantum dot position uncertainty due to fabrication tolerances and stray electric fields. Here, the gradients $b_{\text{deph}}^{\text{dx,dy}}$ are directed parallel to the external applied magnetic field $B_0$ and, in the presence of charge noise, causes dephasing of the qubit\cite{Tokura2006}. $dB_y/dz$ is negligible since the electron wavefunction is strongly confined at the interface, and it is unlikely that the electron position will be shifted in the z-direction due to electric field noise. The field gradient $b_{\text{coup}}$, oriented perpendicular to the external field, enables the spin control through EDSR\cite{Tokura2006}. Finally, $\Delta B_y$ is the field difference between neighbouring qubits and allows for qubit addressability, where we assume an experimentally achievable and commonly used quantum dot pitch of $\SI{100}{\nano \meter}$\cite{Veldhorst2014,Huang2019,Kawakami2014,Yoneda2018}. For optimal operation, $b_{\text{deph}}$ should be low to minimize dephasing, $b_{\text{coup}}$ should be high for fast spin operations, and $\gamma_e \Delta B_y / h$ should be at least $10 \times$ larger than the qubit EDSR linewidth for addressability, where $h$ is Planck's constant. 2D-linecuts for an optimized design are shown in Fig.~\ref{fig:1b}, taken at a quantum dot-to-micromagnet separation ($d_{\text{QDM}}$) of $\SI{120}{\nano \meter}$. 

\begin{figure}
\includegraphics{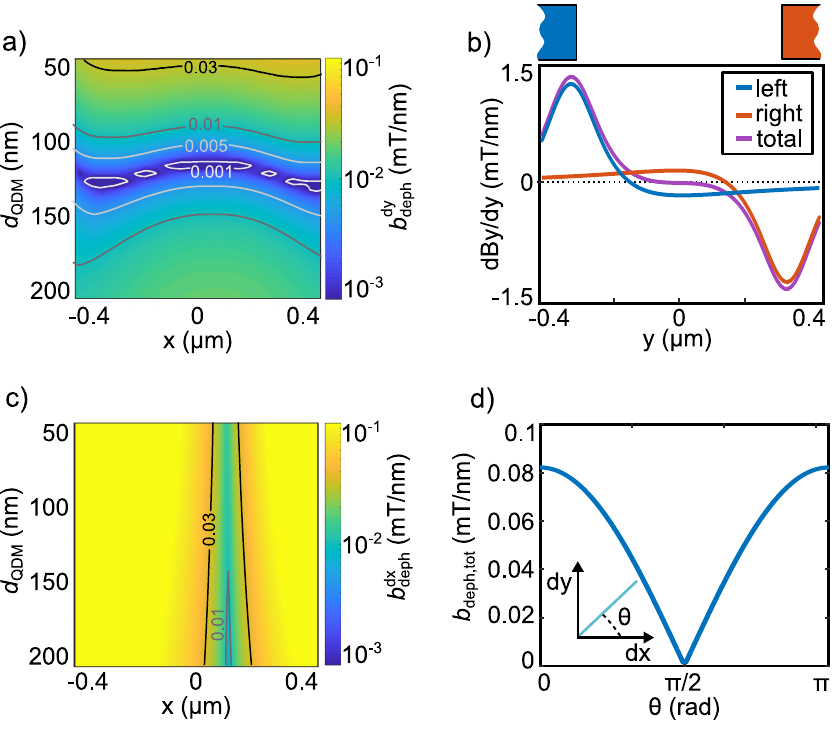}
\caption{\label{fig:2b} a) Dephasing gradient $b_{\text{deph}}^{\text{dy}}$ as a function of the placement of the quantum dots (along the x-direction) and vertical separation $d_{\text{QDM}}$ between the magnets and dots. At an optimal $d_{\text{QDM}}$, $b_{\text{deph}}^{\text{dy}}$ has a minimum where micromagnet induced dephasing effects are strongly reduced. The region where $b_{\text{deph}}^{\text{dy}}$ is minimized stretches out in both x and z dimension. b) This minimum can be explained by the cancellation of the fields generated by the two micromagnet components (blue and red, sum in purple) at a specific $d_{\text{QDM}}$. c) Same as a) but for dephasing gradient $b_{\text{deph}}^{\text{dx}}$. d) Weighted sum of $b_{\text{deph}}^{\text{dx,dy}}$ at $x= \SI{0}{\micro\meter}$ and $d_{\text{QDM}} = \SI{120}{\nano \meter}$ following an angle $\theta$, with $\theta = 0(\pi /2)$ along dx (dy). The dephasing gradient is minimal for $\theta = \pi/2$.}
\end{figure}

The variation of $b_{\text{deph}}^{\text{dy}}$ as a function of $d_{\text{QDM}}$, shown in Fig.~\ref{fig:2b}(a), reveals an optimal $d_{\text{QDM}}$ where $b_{\text{deph}}^{\text{dy}}$ is minimized. The origin of the minimal $b_{\text{deph}}^{\text{dy}}$ is illustrated in Fig.~\ref{fig:2b}(b). The micromagnet components on each side of the gap generate two $b_{\text{deph}}^{\text{dy}}$ that have opposite signs over the central region and hence cancel out. For an optimal $d_{\text{QDM}}$, the total $b_{\text{deph}}^{\text{dy}}$ is almost zero over an extended area, in both x- and y-directions. Although $b_{\text{deph}}^{\text{dx}}$ has no minimum as function of $d_{\text{QDM}}$, Fig.~\ref{fig:2b}(c) shows that there exists a sweetspot along the x-direction where $b_{\text{deph}}^{\text{dx}}$ can be minimized. Eventually, the qubit dephasing rate is determined by the direction of the wavefunction displacement due to charge noise which depends on the details of the device confinement potential and electrical environment. Fig.~\ref{fig:2b}(d) shows that the dephasing gradient is minimized for displacements along the dy-direction. As shown in Fig.~\ref{fig:1b}(c) and Fig.~\ref{fig:2b}(a) the low-dephasing area extends up to a few $\SI{100}{\nano \meter}$ in the qubit plane and $\sim \SI{30}{\nano \meter}$ in the z-direction. At this optimized depth, qubits placed within this area will be highly robust against dephasing arising from charge noise.

\begin{figure}
\includegraphics{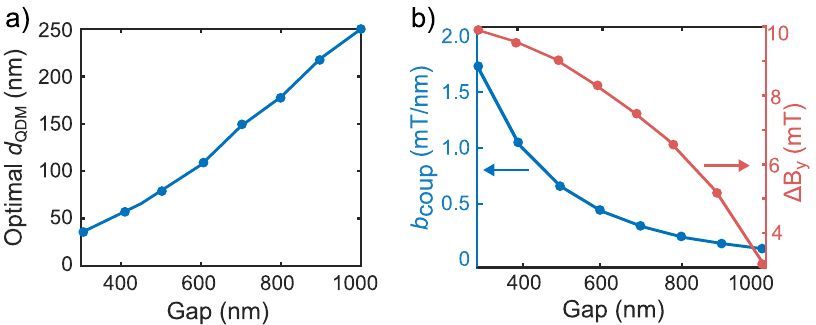}
\caption{\label{fig:3b} a) The optimal depth $d_{\text{QDM}}$ as a function of the gap between the magnets. b) $b_{\text{coup}}$ and $\Delta B_y$ for different gap sizes taken at their respective optimal $d_{\text{QDM}}$, indicating that high speed manipulation and addressability are possible with a minimized dephasing field gradient. }
\end{figure}

So far, we have considered the case where the micromagnet gap size is $\SI{600}{\nano \meter}$. We find for all gap sizes between $\SI{300}{\nano \meter}$ and $\SI{1}{\micro \meter}$ there exists an optimal $d_{\text{QDM}}$, where $b_{\text{deph}}^{\text{dy}}$ is minimal as shown in Fig.~\ref{fig:3b}(a). Furthermore, Fig.~\ref{fig:3b}(b) shows that for each gap size, sufficiently high spin manipulation speed and qubit addressability can be achieved at their respective optimal $d_{\text{QDM}}$. Using the middle point as an example,when the gap is $\SI{600}{\nano \meter}$, $b_{coup} \sim  \SI{0.5}{\milli \tesla / \nano \meter}$ and $\Delta B_y > \SI{5}{\milli \tesla}$. Applying an electron displacement amplitude $ \delta r_d \approx$ $\SI{1}{\nano \meter}$ during EDSR\cite{Kawakami2014}, we can derive a Rabi frequency of $\gamma_e b_{coup} \delta r_d/2h \approx \SI{7}{\mega \hertz}$. The EDSR resonance frequency difference between qubits is given by $\Delta f_0 =  \gamma_e \Delta B_y/h  \approx \SI{140}{\mega \hertz}$, which allows high frequency addressability with little crosstalk.

\begin{figure}
\includegraphics{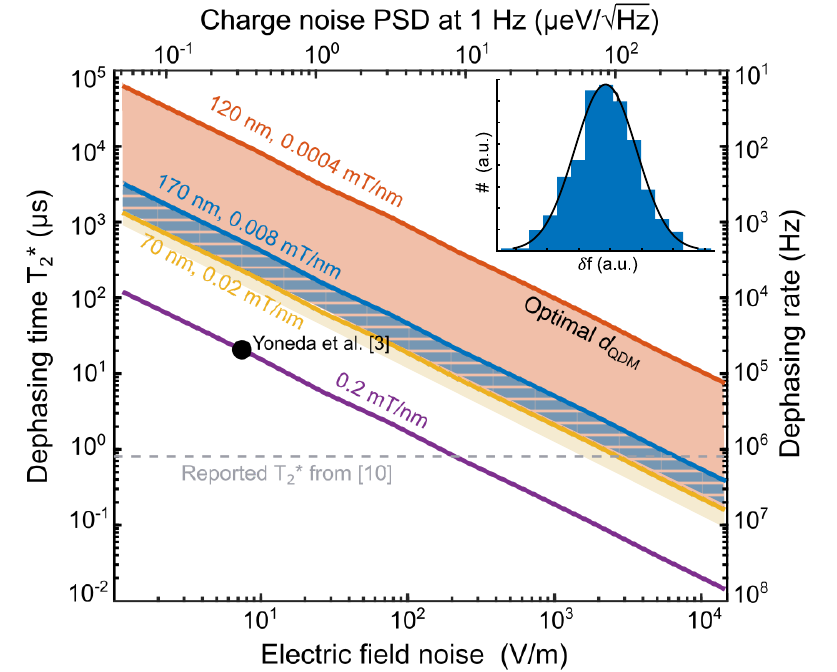}
\caption{\label{fig:4b} Calculated coherence times and dephasing rates due to the presence of $b_{\text{deph}}$, as a function of the electric field noise amplitude and charge noise PSD at 1 Hz. For a $\SI{600}{\nano \meter}$ gap design, three different $d_{\text{QDM}}$ are considered: above, at and below the optimal $d_{\text{QDM}}$ (170, 120 and $\SI{70}{\nano \meter}$ respectively). Bold lines denote the coherence time upper bound assuming displacement along the dy-direction, shaded areas extend to the lower bound for displacements along dx (see also Figure~\ref{fig:2b}d)). An improvement of up to three orders of magnitude can be expected compared to state-of-the-art \cite{Yoneda2015,Kawakami2014}, depending on the displacement direction. Inset shows the histogram of the qubit resonance frequency variation due to the presence of $b_{\text{deph}}$ and charge noise. Dephasing rates are calculated based on the resonance line broadening (Gaussian distribution fit in black). }
\end{figure}

Previous experimental EDSR demonstrations\cite{Kawakami2014,Yoneda2015,Yoneda2018,Noiri2016} made use of a micromagnet design with a gap of $\sim  \SI{300}{\nano \meter}$ and a sub-optimal $d_{\text{QDM}} \sim \SI{250}{\nano \meter}$, and indeed reported dephasing times limited by the detrimental micromagnet gradients and presence of charge noise \cite{Yoneda2018,Struck2020,Takeda2020}. To quantify the importance of a low dephasing field we can relate $b_{\text{deph}}$ to the qubit coherence time $T_{2}^{*}$:
\[
T_{2}^{*} =\frac{h\sqrt{\ln{2}}}{\pi \gamma_e b_{\text{deph}}\delta r_n}
\]
where $\delta r_n$ is the r.m.s electron micromotion induced by the charge noise\cite{Hanson2007}. We make use of a Monte-Carlo simulation to calculate $\delta r_n$ based on the electron wave-function distribution in a symmetrical potential well, in the presence of electric field noise. In Figure \ref{fig:4b}, we compare the dephasing time $T_{2}^{*}$ and dephasing rates ($1/T_{2}^{*}$) due to an electric field noise or charge noise Power Spectral Density (PSD) at $\SI{1}{\hertz}$, for 3 different $d_{\text{QDM}}$. We also add the reported $b_{\text{deph}}$ for Ref. \onlinecite{Yoneda2018,Kawakami2014}. In comparison to these references, an improvement between one and three orders of magnitude in the dephasing rates and dephasing time can be achieved at the optimal $d_{\text{QDM}}$, for displacements along dx or dy, respectively. The ratio $Q$ of qubit dephasing time ($T_{2}^{*}$) and qubit operation time quantifies the fidelity of quantum gate operations in a quantum computer. $Q$ dependents on several factors including the magnitude of charge noise, qubit driving field and is also proportional to the ratio $R = b_{\text{coup}}/b_{\text{deph}}$. High values of $R > 100 $ are achievable at the optimal $d_{\text{QDM}}$ (see supplementary material). 

Finally, to investigate the robustness of the design with respect to the micromagnet topology, we also performed a similar analysis on two other commonly used micromagnet designs (see supplementary material for details). For a design used in single spin quantum electrodynamic demonstrations \cite{Mi2017,Samkharadze2018}, no gradient is present along dx such that low dephasing is ensured regardless of the displacement direction. Optimal gap and depth combinations for qubit operations can be found, and are similar to the design shown in Fig.~\ref{fig:1b}. The optimization robustness against micromagnet topology means it's likely to be useful for other, unexplored micromagnet designs as well.

In summary, we have investigated a widely used micromagnet design for EDSR on semiconductor spin qubits. We demonstrate an optimal and robust quantum dot to magnet separation, where the dephasing field gradient generated by the micromagnet goes to almost zero. Furthermore, we investigated the relation between the micromagnet design parameters and this optimal separation. We show that an optimized micromagnet is achievable for a wide range of fabrication parameters. The designs can be readily extended to one-dimensional arrays, and similar concepts of the symmetrical cancellation shown in Fig~\ref{fig:2b}b) could apply to nanomagnets in two-dimensional arrays as well. However, this would require co-integration of the nanomagnets with gate wiring fanout and is beyond the scope of this work. The presented optimization could lead to an three orders of magnitude increase in dephasing times in comparison to experimental reported state-of-the-art values. As charge noise has been recognized as the limiting factor for spin qubit coherence and fidelity, this optimization will enable spin qubit devices and architectures requiring both long coherence times and high gate fidelities\cite{Vandersypen2017,Li2018,Mohiyaddin2019,Benito2019,Cayao1954}. By optimizing the micromagnet induced dephasing rates, further explorations on other coherence limiting mechanisms can be explored, paving the way towards large-scale spin qubits sytems \cite{Li2018,Vandersypen2017}.

\section*{Supplementary Material}
See Supplementary material for details on the system Hamiltonian, full micromagnet generated fields, influence of the magnetizing field strength, conversion of electric field noise to charge noise power spectrum, and results on the different micromagnet topologies.

\begin{acknowledgments}
The authors acknowledge funding from FWO (Fonds Wetenschappelijk Onderzoek Vlaanderen), grant number 1S60020N (N.D.S.), European Union’s Horizon 2020 Research and Innovation Programme under grant agreement No 951852 (QLSI) and imec’s Quantum Computing IIAP Program.
\end{acknowledgments}

\section*{Data availability}
The data that support the findings of this study are available from the corresponding author upon reasonable request.
\bibliography{aipsamp}

\end{document}